\documentclass[12pt]{article}
\usepackage{paper2e}
\usepackage{mydefs2e}
\usepackage{epsf}

\makeatletter
	\renewcommand{\theequation}{\arabic{equation}}%
	\@addtoreset{equation}{section}%
\makeatother

\begin{document}
\begin{titlepage}

\preprint{UMD-PP-00-82\\UW/PT-00-14}

\begin{center}
\Large\bf
Radion Mediated Supersymmetry Breaking
\end{center}

\author{Z. Chacko}
\address{Department of Physics, Box 351560\\
University of Washington\\
Seattle, Washington 98195\\
{\tt zchacko@fermi.phys.washington.edu}}

\author{Markus A. Luty}
\address{Department of Physics\\
University of Maryland\\
College Park, Maryland 20742\\
{\tt mluty@physics.umd.edu}}

\begin{abstract}
We point out that in supersymmetric theories with extra dimensions, radius
stabilization can give rise to a VEV for the $F$ component of the radius
modulus.
This gives an important contribution to supersymmetry breaking of
fields that propagate in the bulk.
A particularly attractive class of models is obtained if the standard-model
gauge fields propagate in the bulk, while the quark and lepton fields are
localized on a brane.
This leads to gaugino mediated supersymmetry breaking without the need
for singlets in the hidden sector.
We analyze a simple explicit model in which this idea is realized.
\end{abstract}

\date{August 9, 2000}

\end{titlepage}

\noindent
Supergravity (SUGRA) mediated supersymmetry (SUSY) breaking is arguably
the simplest and most natural mechanism for realizing SUSY in nature
\cite{Hidden}.
Because SUGRA couples to everything, it necessarily
connects the observable and hidden sectors, and is therefore a natural
candidate for communicating SUSY breaking to the observable sector.
The main challenges for this class of models are explaining the absence of
flavor-changing squark and slepton mass terms, the absence of
CP violation in soft parameters, and the origin of the $\mu$ term.

The minimal version of SUGRA mediation assumes that all
higher-dimension operators that connect the hidden and observable
sectors are present and suppressed only by powers of $1/M_4$,
where $M_4$ is the 4-dimensional Planck scale.
This gives all required soft SUSY breaking terms (including the $\mu$
term \cite{GM}) of order $m_{3/2}$ from higher-dimension operators of the form
\beq[ops]
\bal
\scr{L}_{\rm eff} &\sim \myint d^4\th\,
\frac{1}{M_4^2} X^\dagger X Q^\dagger Q
+ \myint d^2\th\, \frac{1}{M_4} X \tr W^\al W_\al + \hc
\\
&\qquad
+ \myint d^4\th \left( \frac{1}{M_4} X^\dagger H_u H_d
+ \frac{1}{M_4^2} X^\dagger X H_u H_d + \hc \right)
\eal\eeq
Here, $X$ is a chiral superfield in the hidden sector whose $F$ component
breaks SUSY, $Q$ is a visible sector matter field, $H_u$ and $H_d$ are
Higgs fields, and $W_\al$ is the field strength of the standard model
gauge fields.
Note that the terms that generate the gaugino masses and $\mu$ term require
that $X$ be a singlet.
This scenario is very simple, but it gives no explanation of the absence
of flavor changing scalar masses or CP violation.

A very interesting variation on this scenario is anomaly-mediated SUSY breaking
(AMSB) \cite{RS,GLMR}.
In this scenario, one assumes that the contact terms in \Eq{ops}
connecting the hidden and observable sector are suppressed.
This occurs naturally if the hidden and observable sectors are localized
on different `branes' in a higher-dimensional theory \cite{RS}.
The superconformal anomaly then generates soft masses and $A$ terms of order
$m_{3/2} / 16\pi^2$ \cite{RS,GLMR}.
The minimal version of AMSB predicts negative slepton masses,
but non-minimal visible sectors can make the scenario realistic
\cite{PR,CLMP,KSS}.
Perhaps the most attractive models are those based on the
observation of \Ref{PR}, that thresholds determined by moduli fields can
change the AMSB predictions and make them realistic.
An intriguing possibility is that the GUT threshold plays an important
role \cite{CLPSS}.
The SUSY breaking parameters are naturally real (except possibly the
$\mu$ term), which is sufficient to suppress all CP violation except
for the QCD vacuum angle.
The $\mu$ term requires a separate mechanism; it can be generated in a
relatively simple way as the VEV of a singlet \cite{PR,CLMP}.

Perhaps the simplest version of supergravity mediation is gaugino
mediated SUSY breaking ($\tilde{\rm g}$MSB) \cite{KKS,CLNP}.
This scenario is obtained if the standard-model (or GUT) gauge multiplets
propagate in the bulk of a higher-dimension theory, while the quark and
lepton fields are localized on the visible sector brane.
If the hidden sector is localized on a different brane, the
gauginos can obtain direct contributions to their mass from contact interactions
with the hidden sector of the form (in 5 dimensions)
\beq[contact5]
\De\scr{L}_5 \sim \de(y) \myint d^4\th\, \frac{1}{M_5} X \tr W^\al W_\al + \hc,
\eeq
similar to minimal SUGRA models.
The scalar masses are generated by 1-loop diagrams involving the gauge
fields.%
\footnote{In some versions of this model, the Higgs fields propagate in
the bulk, and therefore obtain soft scalar masses, $\mu$, and $B\mu$
terms directly from contact interactions with the hidden sector \cite{CLNP}.}
The scalar masses are therefore loop suppressed at the compactification
scale, but RG evolution between the compactification scale and the weak scale
can give rise to a realistic superpartner spectrum.
The resulting phenomenology is similar to `no-scale' supergravity models
\cite{noscale}, but it is important to take into account the running above the
GUT scale \cite{SS,SS2} and/or possible Fayet-Iliopoulos terms \cite{CLNP}.
Also, it is important that the geometrical set-up of $\tilde{\rm g}$MSB gives an
understanding of the `no-scale' structure, and why it is stable under radiative
corrections.
This scenario naturally solves the SUSY flavor problem because the gauge couplings
that communicate SUSY breaking to the squarks and sleptons are flavor-blind.
It also solves the SUSY CP problem because the SUSY breaking sector is localized
on a different brane where CP can be a good symmetry.
On the other hand, $\tilde{\rm g}$MSB by itself does not explain the value of
the $\mu$ term.

In $\tilde{\rm g}$MSB, the gaugino masses are conventionally assumed to arise
from a contact terms of the form \Eq{contact5}.
This however requires that the field $X$ that breaks supersymmetry be a singlet.
It is interesting to ask what is the leading SUSY breaking effect in the visible
sector if  the standard model gauge fields are in the bulk but the hidden
sector contains no singlets with large $F$ terms.
A reasonable guess might be anomaly mediation, since the largest competing
contact interaction
\beq
\De\scr{L}_5 \sim \de(y) \myint d^4\th \frac{1}{M_5^2}
X^{\dagger}X ( \tr W^\al W_\al + \hc)
\eeq
is much smaller.
However as we shall show this need not be the case.
There is a large class of theories in which the volume of the extra
dimensions is undetermined in the SUSY limit, and corresponds to a
flat direction called the radius modulus or `radion.'
The low-energy theory below the compactification scale will then contain a
radion chiral superfield $T$.
In order for such a scenario to be realistic, the radius must be stabilized when
SUSY is broken, which corresponds to the radion acquiring a mass.
In this paper, we point out that stabilizing the radius can generate
a VEV for $F_T$.
This gives rise to direct (tree level) SUSY breaking terms for fields
that propagate in the bulk.
Similar effects are well-known in string phenomenology \cite{string}.
If the gauginos propagate in the bulk, this naturally gives rise to a
version of gaugino mediation that we call `radion mediated SUSY breaking'
(RMSB).
We will analyze an explicit model in which the radion is stabilized, which
realizes these ideas.

We work for concreteness in a model with 5
spacetime dimensions, with one dimension compactified on a $S^1/Z_2$
orbifold with radius $r$.
The orbifold projection breaks the $\scr{N} = 2$ SUSY in 5 dimensions down to
$\scr{N} = 1$ in 4 dimensions, and 3-branes may be naturally fixed at the
orbifold fixed points.
We assume for simplicity that the 5-dimensional spacetime is approximately
flat, although the generalization of these results to strongly `warped'
spacetimes \cite{RS1} is interesting.
In SUSY theories, the radion is part of a chiral superfield modulus $T$,
with
\beq[Tdef]
\Re(T) \propto r,
\quad
\Im(T) \propto B_5,
\eeq
where $B_5$ is the $5^{\rm th}$ component of the graviphoton in 5-dimensional SUGRA.%
\footnote{With this choice, the 4-dimensional SUGRA multiplet and $T$ to transform
independently under $\scr{N} = 1$ SUSY.
For an effective field theory derivation of this fact (and \Eq{Tdef}),
see \Ref{LS}.}
Stabilizing the radius requires additional bulk dynamics that is sensitive to $r$.
As we will see in an explicit model, this can give rise to a VEV for $F_T$,
breaking SUSY.

We now work out the effects of $\avg{F_T}$ on a bulk gauge field.
In the 4-dimensional effective lagrangian, the gauge coupling of the gauge
zero mode is given by
\beq[gmatch]
\frac{1}{g_4^2} = \frac{2\pi r}{g_5^2},
\eeq
where $g_5$ is the 5-dimensional gauge coupling
(with mass dimension $-\sfrac{1}{2}$).
To generalize this relation to superspace, note that 
the kinetic term for the gauge zero mode is
\beq[gakin]
\De\scr{L}_4 = \myint d^2\th\, S_4 \tr(W^\al W_\al) + \hc,
\eeq
where $S_4$ is the holomorphic gauge coupling.
The unique SUSY generalization of \Eq{gmatch} is
\beq[holgauge]
S_4 \propto \frac{T}{g_5^2}.
\eeq
From \Eqs{gakin} and \eq{holgauge} we obtain a nonzero gaugino mass
\beq[gauginomass]
m_{\la} = \frac{\avg{F_T}}{2\avg{T}}.
\eeq
We will not consider the analogous results for a bulk hypermultiplet here.
We cannot forbid contact terms between bulk hypermultiplets and
the hidden sector, and therefore the SUSY breaking terms are not completely
determined by $\avg{F_T} / \avg{T}$ for bulk hypermultiplets.

To know whether the contribution \Eq{gauginomass} dominates over other
contributions (such as anomaly mediation), we need to estimate the
quantity $\avg{F_T} / \avg{T}$.
This requires a radius stabilization mechanism.
We will consider the model described in \Ref{LS}.
The additional ingredients required to stabilize the radius are a
$SU(N)$ super Yang--Mills (SYM) sector in the bulk and a
$SU(\tilde{N})$ SYM sector on one of the branes.
Below the compactification scale, the bulk SYM theory becomes a 4-dimensional
SYM sector with a gauge coupling that depends on the radion as in \Eq{holgauge}.
Both SYM sectors give rise to gaugino condensation in the 4-dimensional
theory, generating a dynamical superpotential
\beq
W_{\rm eff} = a e^{-b T} + c.
\eeq
The first term arises from gaugino condensation in the
bulk SYM sector, and the second term in the superpotential arises from the SYM
sector on the brane.
Normalizing $\Re(T) = 3\pi r$ as in \Ref{LS}, we have
\beq
b = \frac{32\pi^2}{3 g_5^2 N}.
\eeq
Using `\naive dimensional analysis' \cite{NDA} we estimate
\beq
a \sim \frac{\La_{\rm UV}^3}{16\pi^2 N},
\qquad
c \sim \frac{\La_{\rm IR}^3}{16\pi^2 \tilde{N}},
\eeq
where $\La_{\rm IR}$ is the scale where the brane SYM sector becomes
strong in the IR, and $\La_{\rm UV}$ is the scale where the 5-dimensional
bulk SYM sector becomes strong in the UV.

In addition to the radius stabilization sector described above,
it is assumed that SUSY is broken on the hidden sector brane.

To describe the complete effective lagrangian, we use the superconformal
formulation of 4-dimensional $\scr{N} = 1$ SUGRA \cite{conformalSUGRA}.
For our results, it will be sufficient to know the
couplings of the scalar auxiliary field of the SUGRA multiplet.
These can be parameterized by the conformal compensator $\phi$, a chiral
superfield with components
\beq
\phi = 1 + \th^2 F_\phi.
\eeq
The couplings of $\phi$ are completely determined by a spurious $U(1)_R$
and dilatation symmetry under which $R(\phi) = \sfrac{2}{3}$, $d(\phi) = 1$.
In a basis where all matter and gauge fields have vanishing $U(1)_R$
charge and dilatation weight, the lagrangian can be written
\beq\eql{LSUGRA}
\scr{L}_4 = \myint d^4\th\, \phi^\dagger \phi f
+ \left( \myint d^2\th\, \phi^3 W + \hc \right),
\eeq
where $W$ is the superpotential, and $f$ is related to the canonically
defined \Kahler potential $K$ by $f = -3 e^{-K/3 M_4^2}$.
Cancelling the cosmological constant after SUSY breaking gives
$\avg{F_\phi} \sim m_{3/2}$.
In this notation, the 4-dimensional effective theory for the model described above is
\beq\bal
\scr{L}_4 &= \myint d^4\th\, \phi^\dagger \phi
\left[ -M_5^3 (T + T^\dagger)
+ f_{\rm vis} + f_{\rm hid} \right]
\\
&\qquad
+ \left[ \myint d^2\th\, \phi^3 \left( a e^{-b T} + c +
W_{\rm vis} + W_{\rm hid} \right) + \hc \right] + \scr{O}(1/M_4^4)
\\
&\qquad
- V_{\rm hid} \times \bigl[ 1 + \hbox{\rm Goldstino\ terms} \bigr].
\eal\eeq
The \Kahler potential has the standard `no scale' form.
$V_{\rm hid}$ is the vacuum energy from SUSY breaking in the hidden sector.
Below the SUSY breaking scale, SUSY will be nonlinearly realized with
additional Goldstino terms that are not relevant here.
We will assume that $a$ and $c$ are real, which corresponds to a discrete
choice of vacua for the gaugino condensates.
Minimizing the resulting potential, we find solutions with $\avg{T}$ real and
$b \avg{T} \gg 1$ for $c \ll a$.
We also obtain
\beq[FTT]
\frac{\avg{F_T}}{\avg{T}} = \frac{2}{b\avg{T}} \avg{F_\phi},
\eeq
which gives SUSY breaking suppressed by a volume factor compared to
$\avg{F_\phi}$.%
\footnote{This suppression was missed in \Ref{LS}, but does not affect
the conclusions of that paper.}
If the standard model gauge multiplet propagates in the bulk, this gives a
contribution to the gaugino masses via \Eq{gauginomass}.

The radion mediated contribution to the gaugino mass from \Eq{FTT} will
be large than the anomaly mediated contribution provided that $b\avg{T}$ is
not too large.
To obtain numerical estimates, we consider the case where the bulk SYM sector
and gravity become strong at the same scale, which gives
$M_5^3 \sim \La_{\rm UV}^3 / 128\pi^3$.
We then solve for $b\avg{T}$ using
\beq[exp]
\frac{\avg{F_\phi}}{M_4} = \frac{a}{M_4^3} b \avg{T} e^{-b \avg{T}}.
\eeq
From $\avg{F_\phi} \sim m_{3/2} \sim 10$--$100\TeV$, we obtain
$b\avg{T} \simeq 32$.
This estimate is dominated by the exponent in \Eq{exp}, and is therefore
insensitive to the estimates used above.
We then obtain
\beq
\frac{\avg{F_T}}{\avg{T}} \sim \frac{1}{16} \avg{F_\phi},
\qquad
r \La_{\rm UV} \sim 25.
\eeq
Note that the radius is sufficiently small that the standard-model
gauge coupling is perturbative up to the scale $\La_{\rm UV}$.
On the other hand, the radius is large enough to suppress the
exchange of massive states (\eg excited string modes or extended objects)
between the hidden and visible sector branes.
These are potentially dangerous because they can give rise to
contact terms of the form
\beq
\De \scr{L}_4 \sim \myint d^4\th\, \frac{e^{-M r}}{M^2} X^\dagger X Q^\dagger Q.
\eeq
where $M$ is the mass of the heavy states.
Since we expect $M \sim \La_{\rm UV}$, this is sufficient to suppress
FCNC's.

We see that in this model the radion-mediated contribution to the gaugino mass is
about an order of magnitude smaller than $m_{3/2}$.
However, it is still larger than the anomaly mediated contribution
$\De m_\la \sim m_{3/2} / 16\pi^2$.
If the compactification scale is above the GUT scale, it may appear that large
multiplicity factors in loops above the GUT scale can make the anomaly-mediated
contribution to the gaugino mass comparable to (or larger than) the
radion-mediated contribution.
However, this generally does not happen.
Note that at the compactification scale, the gaugino mass is (up to loop matching
corrections) the sum of the AMSB and RMSB contributions.
(Cross terms arising from \eg radion dependent SUSY breaking in the regulators
is loop suppressed compared to the direct RMSB contribution.)
RG evolution from the compactification scale down to the GUT scale corrects
this by factors that are loop suppressed compared to the gaugino
mass at the compactification scale.
Even if the AMSB contribution dominates above the GUT scale, the RG is dominated
by the AMSB trajectory, and the the gaugino mass at the GUT scale is dominantly
AMSB, with an additive correction equal to the RMSB contribution.
If the GUT scale is a SUSY threshold (unlike the model of \Ref{CLPSS}) there
are large loop matching contributions that will reduce the AMSB contributions to the
corresponding MSSM values below the GUT scale.
One is therefore left with the RMSB contribution with a small AMSB correction.
A brief calculation shows that just below the GUT scale the ratio of the
magnitudes of the two contributions is approximately 3:1 for the gluino, 10:1
for the wino and 3:2 for the bino, and the total gaugino mass is the sum (or
difference) of the two contributions.
To avoid a small bino mass, we require that the AMSB and RMSB contributions
have the same sign, which is obtained for an appropriate choice of sign for $a$
in the model above.
The sizable correction to the bino mass in this model will show up as
a correction to gaugino unification in this model.

A similar argument shows that the AMSB contributions to the scalar
masses are small below the GUT scale.

We conclude that gaugino mediation arises naturally in this model.
The corrections to the bino mass will increase the slepton masses, which
may have important phenomenological implications.
We will leave this for future work.

In conclusion, we have shown that that radius stabilization can give rise
to a VEV for the $F$ component of the radion modulus, which can give an
important contribution to SUSY breaking of bulk fields.
If the standard model (or GUT) gauginos propagate in the bulk, while the
other fields are localized on branes, we naturally obtain gaugino
mediated SUSY breaking without the need for singlets in the hidden sector.
We have analyzed a specific radius stabilization mechanism to illustrate
these ideas and give a completely realistic model.
We hope that these ideas will show the way to even simpler realistic
models, perhaps even in the context of string theory.

\emph{Note added:}
While this paper was being completed, we received \Ref{KKSUSY}, which
considers related ideas.

\section*{Acknowledgments}
We thank R. Sundrum for discussions.
Z.C. was supported by the DOE under contract DE-FG03-96-ER40956.
M.A.L. was supported by the NSF under grant PHY-98-02551.


\end{document}